\documentclass{aa}  
\usepackage{graphicx}
\usepackage[varg]{txfonts}
\usepackage{natbib}
\usepackage{graphicx}
\usepackage{gensymb}
\usepackage{url}
\usepackage{amssymb,amsmath}
\begin{document} 

\title{
Galactic Bulge Population II Cepheids in the VVV Survey: Period-Luminosity Relations and a Distance to the Galactic Center}

\author{A. Bhardwaj\inst{1,2}
          \and 
          M. Rejkuba\inst{1,3} 
	   \and
          D. Minniti\inst{4,5,6} 
	   \and
          F. Surot\inst{1} 
	   \and
          E. Valenti\inst{1} 
	   \and
          M. Zoccali\inst{7,5}
	   \and
          O. A. Gonzalez\inst{8}
	   \and \\
          M. Romaniello\inst{1,3}
	   \and
          S. M. Kanbur\inst{9}    
          \and 
          H. P. Singh\inst{2}
           }
\institute{European Southern Observatory, Karl-Schwarzschild-Stra\ss e 2, 85748, Garching, Germany\\
   	   \email{anupam.bhardwajj@gmail.com; abhardwaj@eso.org}
         \and
             Department of Physics and Astrophysics, University of Delhi,  Delhi-110007, India 
         \and   
             Excellence Cluster Universe, Boltzmann-Str. 2, D85748 Garching bei M$\ddot{\mathrm{u}}$nchen, Germany
         \and   
	     Departamento de F\'isica, Facultad de Ciencias Exactas, Universidad Andr\'es Bello Av. Fernandez Concha 700, 
             Las Condes, Santiago, Chile
         \and   
             Instituto Milenio de Astrofisica, Santiago, Chile.
         \and   
             Vatican  Observatory,   V00120  Vatican  City  State, Italy.
       	\and
             Pontificia Universidad Cat\'olica de Chile, Instituto de Astrof\'isica, Av. Vicu\~na Mackenna 4860, Santiago, Chile
        \and
             UK Astronomy Technology Centre, Royal Observatory, Blackford Hill, Edinburgh EH9 3HJ, UK
        \and
	     State University of New York, Oswego, NY 13126, USA
             }

\date{Received xxxx; accepted xxxx}

  \abstract
   {Multiple stellar populations of different ages and metallicities reside in the Galactic bulge tracing its structure and providing clues for its formation and evolution.}
{We present the near-infrared observations of population II Cepheids in the Galactic bulge from VVV survey. The $JHK_s$ photometry together with optical data from OGLE survey provide an independent estimate of the distance to the Galactic center. The old, metal-poor and low-mass population II Cepheids are also investigated as useful tracers for the structure of the Galactic bulge.}
{We identify 340 population II Cepheids in the VVV survey Galactic bulge catalogue based on their match with OGLE-III Catalogue. The single-epoch $JH$ and multi-epoch $K_s$ observations complement the accurate periods and optical $(VI)$ mean-magnitudes from OGLE. The sample consisting of BL Herculis and W Virginis subtypes is used to derive period-luminosity relations after correcting mean-magnitudes for the extinction. Our $K_s$-band period-luminosity relation, $K_s = -2.189(0.056)~[\log(P) - 1] + 11.187(0.032)$, is consistent with published work for BL Herculis and W Virginis variables in the Large Magellanic Cloud.}
{We present a combined OGLE III and VVV catalogue with periods, classification, mean magnitudes and extinction for 264 Galactic bulge population II Cepheids having good-quality $K_s$-band light curves. The absolute magnitudes for population II Cepheids and RR Lyraes calibrated using {\it Gaia} and {\it Hubble Space Telescope} parallaxes, together with calibrated magnitudes for Large Magellanic Cloud population II Cepheids, are used to obtain a distance to the Galactic center,  $R_0=8.34\pm0.03{\mathrm{(stat.)}}\pm0.41{\mathrm{(syst.)}}$, which changes by $^{+ 0.05}_{-0.25}$ with different extinction laws. While noting the limitation of small number statistics, we find that the present sample of population II Cepheids in the Galactic bulge shows a nearly spheroidal spatial distribution, similar to metal-poor RR Lyrae variables. We do not find evidence of the inclined bar as traced by the metal-rich red-clump stars.}
{Population II Cepheid and RR Lyrae variables follow similar period-luminosity relations and trace the same metal-poor old population in the Galactic bulge. The number density for population II Cepheids is more limited as compared to abundant RR Lyraes but they are bright and exhibit a wide range in period that provides a robust period-luminosity relation for an accurate estimate of the distance to the Galactic center.}

\titlerunning{Type II Cepheids in the Galactic Bulge}
\maketitle
%

\section{Introduction}
\label{sec:intro}
Population II Cepheids are low-mass, metal-poor stars that are found in globular clusters, Galactic disk and bulge populations \citep{wallerstein2002, sandage2006}. These Type II Cepheid (T2C) variables are more than a magnitude fainter than Classical or Type I Cepheids with similar periods and follow a shallower Period-Luminosity relation \citep[PLR or ``Leavitt Law'',][]{leavitt2012}. T2Cs reside in the instability strip just above (brighter) RR Lyrae (RRL) variables and are subdivided into three subclasses, BL Herculis (BLH), W Virginis (WVR)  and RV Tauris (RVT). These subclasses represent different evolutionary states with short-period BLH moving from blue horizontal branch to asymptotic giant branch (AGB), intermediate period WVR stars undergo helium shell flashes and make temporary excursions from the AGB into the instability strip, while long period RVT suggest post-AGB evolution \citep{wallerstein2002}. The characteristic light curves for the subclasses of T2Cs are different and the PLRs exhibit a greater dispersion and non-linearity at optical wavelengths \citep[][and references within]{nemec1994, alcock1998, kubiak2003, majaess2009, schmidt2009}, thus, limiting their use as primary distance indicators. At near-infrared (NIR) wavelengths, T2Cs and RRLs follow similar PLRs \citep{sollima2006, matsunaga2006, ripepi2015, cpapir4} and T2Cs being relatively bright variables, can be used to obtain robust distances. These variables provide an independent method to determine the distance to the Galactic center and to trace the structure of the old stellar population in the Galactic bulge. For example, \citet{gmat2008} estimated a T2C and RRL based distance to the Galactic center, $R_0=7.94\pm0.37$~kpc.

The optical light curves of T2Cs in the Galactic bulge and the Magellanic Clouds (MC) are provided by the third phase of the Optical Gravitational Lensing Experiment (OGLE-III) survey \citep{soszynski2008a, soszynski2010, soszynski2011}. The NIR studies of these variables in the MC were carried out by \citet{matsunaga2009, matsunaga2011}, \citet{ciech2010}, \citet{ripepi2015} and \citet{cpapir4}. T2Cs in the Galactic globular clusters at $JHK_s$ wavelengths were observed by \citet{matsunaga2006}. These authors discussed the T2C PLRs and their distance scale applications at NIR wavelengths. The time-series NIR observations for T2Cs in the Galactic bulge have been limited to a sample of 39 Cepheids observed by \citet{gmat2008}. 

The VISTA Variables in the V\'ia L\'actea (VVV) survey \citep{minnitivvv} has provided a large amount of variable star data to probe the 3-D structure of the Galactic bulge \citep[][and reference therein]{wegg2013, gonzalez2013, valenti2016, zoccali2016}. Thanks to high-resolution spectroscopic investigations of a sizeable sample of bulge giants \citep[][and references therein]{zoccali2008, Hill2011, Ness2013, Rojas2014, Gonzalez2015a} it is now widely accepted that the bulge metallicity distribution is broad (i.e. $-1\leq [Fe/H] \leq +0.5$~dex), bimodal, and with two peaks few dex below and above the solar value. The observed metallicity gradient as a function of the height from the Galactic plane is due to the change of the relative fraction of the metal-rich and metal-poor components along the line-of-sight. On the other hands, the vast majority of the photometric studies aimed at dating the bulge stars \citep{ortolani1995, Kuijken2002, zoccali2003, Sahu2006, clarkson2008, clarkson2011, valenti2013} revealed predominantly old stellar population. However, according to spectroscopic microlensing follow-up \citep[][and references therein]{bensby2017} a smaller but significant fraction of young and intermediate age population may also be present. The old metal-poor tracers of the bulge show an axisymmetric and spheroidal distribution \citep[RRLs, Miras, Red-clump giants, ][]{dekeny2013, catchpole2016, gran2016, zoccali2017}, or a triaxial ellipsoidal distribution \citep[RRLs,][]{piet2015}. The metal-rich red-clump giants trace the X-shaped bar in the bulge \citep{mcwilliam2010, nataf2010, saito2011, gonzalez2015, zoccali2017}. 

Recently, \citet{minniti2016} discovered a dozen fundamental-mode RRL variables in the vicinity of the Galactic Center. T2Cs are less abundant but brighter than RRL and therefore, are easy to find in highly extincted regions in the bulge. A search for new T2Cs in the whole VVV bulge area will be presented elsewhere. For the present analysis, we use the sample that has optical counterparts in the OGLE catalog. 

The paper is structured as follows: We discuss photometry of T2Cs from VVV survey in Section~\ref{sec:data}. We derive PLRs at $JHK_s$ wavelengths and compare our results with published work in \S3. We determine a distance to the Galactic center using calibrated absolute magnitudes for T2Cs in \S4. In Section~\ref{sec:spatial}, we discuss the spatial distribution of T2Cs in the Galactic bulge and compare our results with RRLs and red-clump stars. We summarize our results in Section~\ref{sec:discuss}.

\section{The Data}
\label{sec:data}

We present near-infrared photometry of T2Cs from VVV survey \citep{minnitivvv, saitovvv} latest DR4 catalog (Hempel et.\ al, in prep.). We perform a cross-match of positions for OGLE-III bulge T2C sample to this catalog within $1''$ search radius to identify 340 Cepheids. The median separation between OGLE and VVV sources is $0.08''$ with a standard deviation of $0.15''$ and more than $95\%$ of the matched sources have a separation of $\lesssim0.5''$. We adopt the classification based on $I$-band light curves from OGLE-III and there are 147 BLH, 123 WVR and 70 RVT type variables with $JHK_s$ observations. 

\begin{figure}
\begin{center}
\includegraphics[width=0.5\textwidth,keepaspectratio]{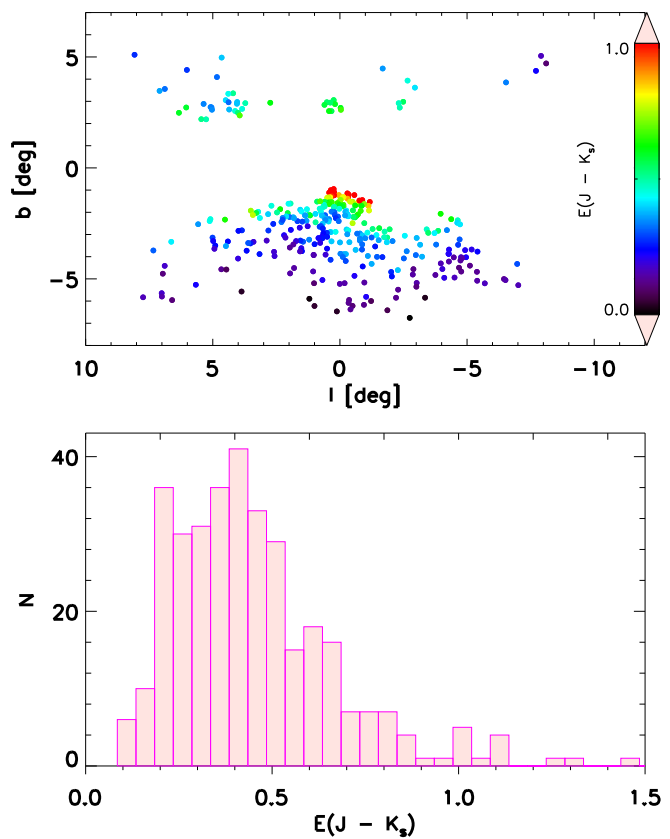}
\caption{Top panel : The spatial distribution and the $E(J-K_s)$ color excess for Type II Cepheids in the VVV survey. Bottom panel : Histogram of the $E(J-K_s)$ color excess. Using reddening law of \citet{nishiyama2009}, the extinction in $K_s$-band amounts to, $A_{K_{s}} = 0.528E(J-K_s)$.}
\label{fig:ejk_vvv}
\end{center}
\end{figure}

The VVV $JH$-band magnitudes are single epoch observations while the $K_s$-band has multi-epoch data. The number of observations in $K_s$ varies from $\sim10$ to $\sim100$ depending on the location and brightness of the T2C, and on average there are $\sim50$ epochs per light curve. The typical apparent magnitudes for T2Cs in $K_s$ ranges from $\sim15$ to $\sim10$~mag. The period ($P$), time of maximum brightness in $I$-band and the optical ($VI$) mean-magnitudes for the matched T2Cs are taken from OGLE-III catalog \citep{soszynski2011}. The $E(J-K_s)$ color excess for T2Cs in the bulge is obtained using the extinction maps of \citet{gonzalez2011, gonzalez2012}. Fig.~\ref{fig:ejk_vvv} displays the spatial distribution and $E(J-K_s)$ color excess for all 340 T2Cs and the histogram of color excess is shown in the bottom panel.

\begin{figure}
\begin{center}
\includegraphics[width=0.5\textwidth,keepaspectratio]{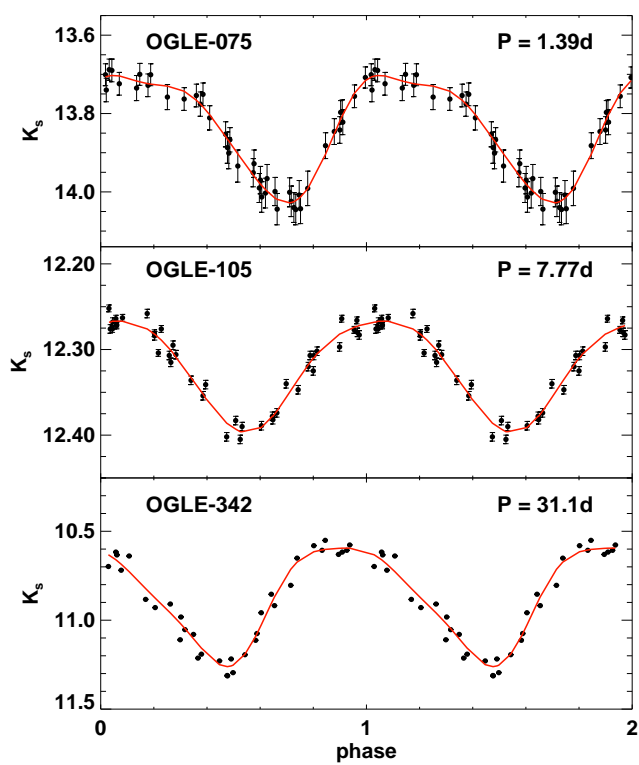}
\caption{The representative light curves in $K_s$-band for Type II Cepheids in the VVV survey. Top/middle/bottom panels display the BLH/WVR/RVT type variables. The solid red line is the Fourier fit to the light curve. Star IDs and periods are provided at the top of each panel.}
\label{fig:lc_vvv}
\end{center}
\end{figure}

\begin{table*}
\begin{center}
\caption{Properties of Type II Cepheids in the Galactic bulge. \label{table:t2c}}
\scalebox{0.95}{
\begin{tabular}{ccccccccccccc}
\hline
\hline
VVV ID  &  OGLE ID&   P & Class&         \multicolumn{5}{c}{Mean magnitudes}&   \multicolumn{3}{c}{$\sigma$}&       $E_{JK_s}$\\
	&         &(days)&      &V&      I&      J&      H&      $K_S$&       J&      H&      $K_S$&          \\
\hline
\hline
515601356315& OGLE-002&      2.268& BLH&     15.188&     13.909&     13.116&     12.766&     12.521&      0.072&      0.072&      0.072&      0.173\\
515601679485& OGLE-003&      1.484& BLH&     16.519&     15.061&     14.240&     13.855&     13.387&      0.104&      0.104&      0.104&      0.290\\
515594023082& OGLE-005&      2.008& BLH&     18.666&     16.842&     15.845&     15.354&     14.788&      0.152&      0.153&      0.150&      0.388\\
515520862858& OGLE-008&      1.183& BLH&     17.765&     15.970&        ---&     14.205&     13.935&        ---&      0.060&      0.058&      0.410\\
515555436341& OGLE-009&      1.896& BLH&     17.630&     15.608&     13.887&     13.343&     13.134&      0.098&      0.098&      0.098&      0.461\\
515534145302& OGLE-013&      1.517& BLH&     18.525&     16.193&     14.550&     13.951&     13.676&      0.117&      0.117&      0.116&      0.588\\
515543870338& OGLE-014&      1.239& BLH&     15.376&     13.590&     12.306&     11.905&     11.613&      0.003&      0.004&      0.092&      0.534\\
515490128302& OGLE-015&      1.279& BLH&     18.087&     15.907&     14.291&     13.712&     13.438&      0.008&      0.010&      0.111&      0.591\\
515522173203& OGLE-017&      1.098& BLH&     18.533&     16.251&     14.371&     13.850&     13.669&      0.091&      0.092&      0.090&      0.636\\
515490198825& OGLE-018&      1.620& BLH&     18.072&     15.990&     14.187&     13.630&     13.339&      0.113&      0.113&      0.113&      0.580\\
\hline
\end{tabular}}
\end{center}
{\footnotesize {{\bf Notes:} The OGLE ID, period, subtype and optical mean magnitudes are taken from OGLE-III \citep{soszynski2011}.
$E_{JK_s}~:~E(J-K_s)$ values are taken from extinction maps of \citet{gonzalez2011}. The first ten lines of the table are shown here and the entire table is available online as supplemental material.}}
\end{table*}

In order to select the best-quality light curves, we only considered stars with a minimum of 20 $K_s$ measurements. We use periods and time of maximum brightness from OGLE to phase these light curves and apply a fifth order Fourier-series fit \citep[see,][]{bhardwaj2015} to determine the peak-to-peak amplitudes ($A$) and the standard deviation ($\sigma$) of the fit. We limit our final sample to the light curves for which, $\sigma/A<1/20$. This provides an estimate of the impact that photometric uncertainties have on the shape of the light-curves.  
Finally, we also remove 4 noisy light curves with very low-amplitudes ($A < 0.08$~mag). Following these selection criteria, we are left with a final sample of 264 T2Cs. Fig.~\ref{fig:lc_vvv} displays the representative light curves of BLH, WVR and RVT stars from our final sample of good-quality light-curves. 

The saturation limit for $K_s$-band in VVV survey is $\sim11$~mag \citep[see, Figure 2 in][]{gonzalez2011} and the deviation from 2MASS occurs around 12~mag in $JH$. While the higher extinction in the bulge can make these sources fainter, some of the long-period T2Cs may indeed be affected by non-linearity and saturation. We note that most of the rejected light-curves belong to long-period bright RVT stars and therefore, this subclass will not be used for P-L relations and distance estimates.

\section{Period-Luminosity relations for Type II Cepheids in the Bulge}
\label{sec:vvv}

The mean-magnitudes in $K_s$-band are estimated from the multi-epoch VVV data by fitting templates from \citet{cpapir4} for T2Cs. For each single-epoch $JH$ measurement, we estimate the phase using the time of maximum brightness in $I$-band from OGLE survey. We use $IK_s$-band templates and apply phase correction to $J$ and $H$-band magnitudes to derive their mean values. The mean properties of T2Cs in our final sample from OGLE and VVV are listed in Table~\ref{table:t2c}. 

Several studies addressed the reddening law towards the Galactic center \citep[e.g.][]{nishiyama2006, nishiyama2009, gonzalez2012, nataf2016, majaess2016} and its impact on the distance estimates will be discussed in the next section. For now, we adopt the \citet{nishiyama2009} reddening law and total-to-selective absorption ratios, $R_J = 1.526$, $R_H = 0.855$ and $R_K = 0.528$ corresponding to $E(J-K_s)$, to apply extinction corrections. The $E(J-K_s)$ value is greater than 1 for only 11 Cepheids in our sample. We also include reddening independent optical Wesenheit, $W_{V,I} = I - R^V_I(V-I)$, where the absorption-ratio, $R^V_I=1.08$, is adopted from \citet{soszynski2011}. The value of $R^V_I$ changes significantly depending on the choice of reddening law but we follow OGLE relation for a relative comparison under the assumption that it exhibits the least scatter in the optical Wesenheit. We assume that all T2Cs are at the same distance and fit a PLR in the following form:

\begin{equation}\label{eq:plr}
m^0_{\lambda} = a_{\lambda}[\log(P) - 1] +b_{\lambda},
\end{equation}

\begin{figure}
\begin{center}
\includegraphics[width=0.49\textwidth,keepaspectratio]{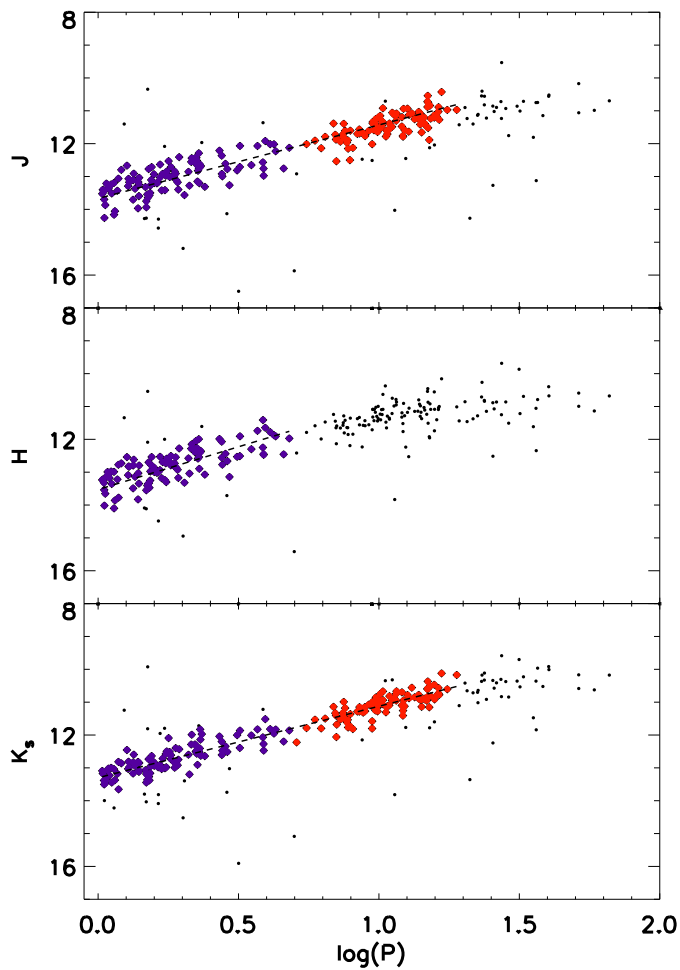}
\caption{Extinction-corrected near-infrared PLRs for Type II Cepheids in the Galactic bulge. The violet and red symbols represent BLH and WVR stars, respectively. The dashed line represents a single regression line fit over colored symbols and small symbols represent the Type II Cepheids displaying evidence of approaching saturation in the long period range and $2\sigma$ outliers in the short-period range. $H$-band data for WVR stars are also close to saturation limit and therefore we do not include them in linear regression fit. }
\label{fig:pl_vvv}
\end{center}
\end{figure}

\begin{table}
\begin{center}
\caption{Galactic bulge Type II Cepheid PLRs.}
\label{table:gb_pl}
\scalebox{0.93}{
\begin{tabular}{cccccc}
\hline
\hline
Band & Type & Slope & Intercept & $\sigma$ &  N   \\
\hline
     $W_{V,I}$&      B+W&    -2.294$\pm$0.055     &    11.511$\pm$0.030     &     0.342&         212\\
             J&      BLH&    -2.387$\pm$0.164     &    11.393$\pm$0.132     &     0.347&         106\\
             J&      WVI&    -2.037$\pm$0.096     &    11.476$\pm$0.012     &     0.242&          93\\
             J&      B+W&    -2.240$\pm$0.031     &    11.495$\pm$0.009     &     0.316&         203\\
             H&      BLH&    -2.591$\pm$0.163     &    11.019$\pm$0.130     &     0.353&         104\\
         $K_s$&      BLH&    -2.362$\pm$0.170     &    11.071$\pm$0.133     &     0.294&         108\\
         $K_s$&      WVI&    -2.373$\pm$0.272     &    11.194$\pm$0.034     &     0.194&          95\\
         {\bf $K_s$}&     {\bf  B+W}&    {\bf -2.189$\pm$0.056}     &    {\bf 11.187$\pm$0.032}     &     {\bf 0.234}&         {\bf 201}\\
\hline
\end{tabular}}
\end{center}
{\footnotesize {\bf Notes:} B+W : BL Herculis and W Virginis.}
\end{table}

\noindent where $m^0_\lambda$ is the extinction-corrected mean-magnitude from VVV survey and $\lambda$ represents the $JHK_s$ wavelengths. 
The coefficient $a$ is the slope and $b$ is the zeropoint at $P=10$~d. We fit this equation to the BLH and WVR classes separately and also to the combined sample of BLH+WVR and iteratively remove $2\sigma$ outliers in all cases. The adopted threshold provides a stronger constraint on the slopes and zeropoints of the PLRs and will be followed throughout the paper. We note that a higher sigma-clipping does not change the PLRs significantly and the number of stars and the dispersion increases marginally. Fig.~\ref{fig:pl_vvv} displays the PLRs for T2Cs in the Galactic bulge. We observe a flattening of the $H$-band PLR for WVR variables. After looking at several $H$-band images, we find that these stars show evidence of approaching saturation. Therefore, we will use only BLH type variables in the $H$-band for the present analysis. The slopes and intercepts of PLRs are listed in Table~\ref{table:gb_pl}. The dispersion in PLR is presumably dominated by the depth of the bulge with possible contribution due to the inner Galactic bar \citep{nishiyama2005, gonzalez2011a}.

\subsection{Comparison with published PLRs}

We compare the Galactic bulge PLRs for T2Cs with published work in the Galactic globular clusters, the bulge and the LMC from \citet{matsunaga2006}, \citet{gmat2008} and \citet{cpapir4}. We note that \citet{cpapir4} derived new PLRs for T2Cs in the LMC and found the slopes and intercepts to be consistent with previous results. Therefore, we only consider PLRs in the LMC from \citet{cpapir4}. We also compare the optical Wesenheit in the bulge with LMC, where total-to-selective absorption ratio, $R^V_I=1.55$, is taken from \citet{soszynski2010} for LMC T2Cs. We note that optical Wesenheits for T2Cs in the LMC and bulge are adopted only for a relative comparison and these relations will not be used for distance estimates. We will use standard t-test to compare the PLRs, given the uncertainties in the slopes and the {\it rms} of the relation under consideration. The details of the test-statistics is discussed in \citet{bhardwaj2016a}. In brief, the null hypothesis i.e. the two slopes are equal, is rejected if the probability  $p(t)\!<\!0.05$. 

\begin{table*}
\begin{minipage}{1.0\hsize}
\begin{center}
\caption{Comparison of Galactic bulge Type II Cepheids with PLRs from literature.}
\label{table:comp_pl}
\begin{tabular}{ccccccccc}
\hline
\hline
Band & Slope & $\sigma$ &  N & Host & Type & Source& |T| & $p(t)$   \\
\hline
     $W_{V,I}$&    -2.294$\pm$0.055     &      0.342&     212     &GB&B+W&TW& ---& ---\\
              &    -2.677$\pm$0.052     &      0.176&     186    &LMC&all&TW&      4.265&      0.000\\
              &    -2.508$\pm$0.074     &      0.111&     133    &LMC&B+W&TW&      1.171&      0.242\\
         $K_s$&    -2.189$\pm$0.056     &      0.234&     201     &GB&B+W&TW& ---& ---\\
              &    -2.240$\pm$0.140     &      0.410&      39    &GB&all&G08&      0.462&      0.644\\
              &    -2.395$\pm$0.027     &      0.228&     167   &LMC&all&B17&      3.621&      0.000\\
              &    -2.232$\pm$0.037     &      0.180&     119   &LMC&B+W&B17&      0.648&      0.517\\
              &    -2.413$\pm$0.053     &      0.150&      43   &GGC&all&M06&      2.309&      0.022\\
              &    -2.425$\pm$0.295     &      0.075&       9 &MW&all$^a$&B17&      0.243&      0.808\\
\hline
\end{tabular}
\end{center}
{\footnotesize {\bf Notes:} B+W : BL Herculis and W Virginis; $^a$ represents absolute calibration of PLRs based on parallaxes for T2Cs and RRLs. Source column represents - TW : This work; G08 : \citet{gmat2008}; B17 : \citet{cpapir4}; M06 : \citet{matsunaga2006} PLR with updated mean-magnitudes from \citet{cpapir4}. Last two columns represent the observed value of the t-statistic (|T|) and the probability, $p(t)$, of acceptance of the null hypothesis.}
\end{minipage}
\end{table*}

\begin{figure}
\begin{center}
\includegraphics[width=0.49\textwidth,keepaspectratio]{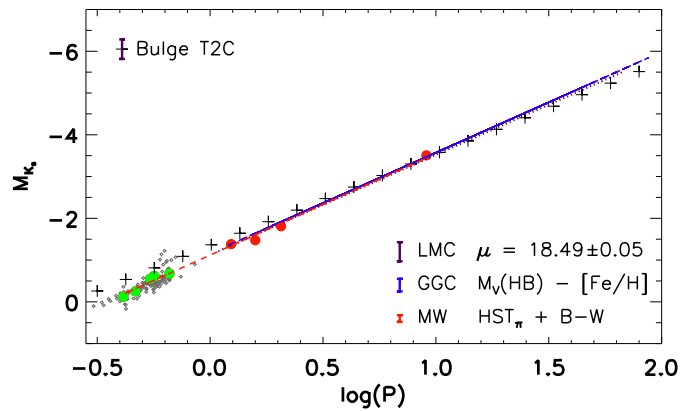}
\caption{Calibrated PLRs for Type II Cepheids in the LMC, Galactic globular clusters (GGC) and the Milky Way (solar neighbourhood). The red and green circles represent absolute magnitudes for T2Cs and RRLs with parallaxes  in the Milky Way and dashed red line is the best-fit linear regression. Plus symbols display PLR for VVV T2Cs with a zero-point offset with respect to calibrated magnitudes. The error bars represent $1\sigma$ dispersion in the P-L relation in each galaxy. Grey dots display absolute magnitudes for RRLs in the LMC and globular clusters.}
\label{fig:pl_cal}
\end{center}
\end{figure}

The results of t-test statistics are listed in Table~\ref{table:comp_pl}. We find that the $W_{V,I}$ Wesenheits and $K_s$-band PLRs in the Galactic bulge and the LMC are consistent, if we consider only BLH+WVR variables. The slope of the $K_s$-band PLR is consistent with the result of \citet{gmat2008} for the bulge, while there is a marginal but statistically significant difference in slopes with Galactic globular clusters. We also compare the $K_s$-band PLRs with the calibrated PLRs for T2Cs and RRLs using {\it Gaia} and {\it HST} parallaxes from \citet{cpapir4}. The PLRs are consistent, given the large uncertainties in the slope of the calibrated PLR. 

Fig.~\ref{fig:pl_cal} displays the calibrated PLRs in the LMC (violet), Galactic globular clusters (blue) and the MW (red). We adopt the late-type eclipsing binary distance of $18.493\pm0.047$~mag \citep{piet13} to calibrate LMC PLR. The two T2Cs and five RRLs in the solar neighbourhood are calibrated with available trigonometric parallaxes from {\it Gaia} and {\it HST} \citep[see,][]{cpapir4}. We also include two T2Cs with pulsation parallaxes from \citet{feast2008}. We note that the entire period-range $K_s$-band PLRs in the LMC, Galactic globular cluster and in the MW have nearly the same slopes and zero-points. This provides an additional evidence that $K_s$-band PLRs are less sensitive to the metallicity and extinction and can be used to obtain accurate distance estimates. The $K_s$ mean-magnitudes for RRLs are also included to extend the PLRs followed by T2Cs (see, grey symbols in Fig.~\ref{fig:pl_cal}). The near-infrared photometry of RRLs in the LMC is taken from \citet{borissova2009,muraveva2015}, while the globular clusters RRLs data is adopted from M92 and M4 \citep{del2005, stetson2014}. This further confirms the consistency between the distance scale for T2Cs and RRLs as discussed in previous papers \citep{sollima2006, matsunaga2006, ripepi2015, cpapir4}.

\section{Distance to the Galactic Center}

\begin{figure}
\begin{center}
\includegraphics[width=0.5\textwidth,keepaspectratio]{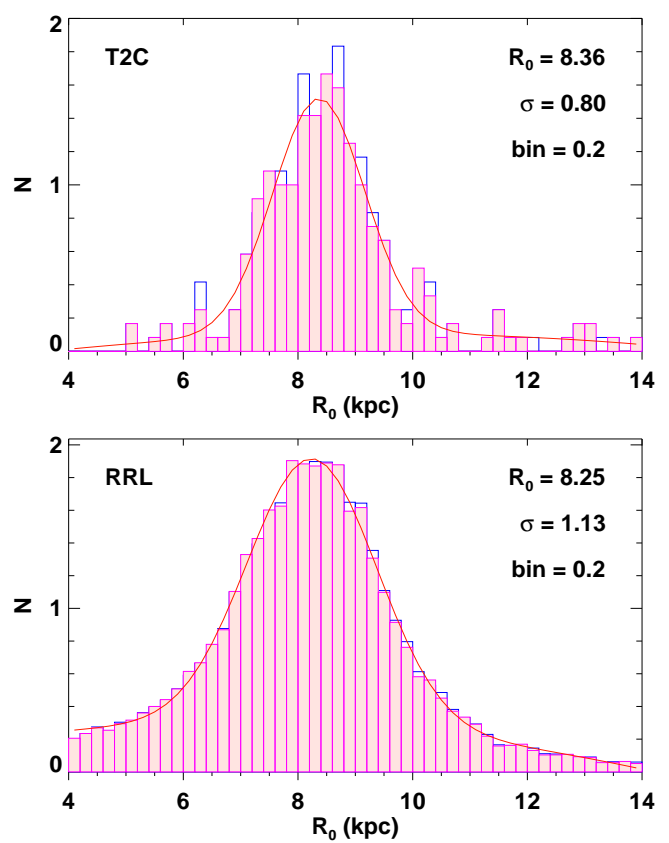}
\caption{Top panel : Histogram of the projected distances of BLH+WVR T2Cs. The blue/magenta bars represent the distance distribution before and after the geometric corrections. Bottom panel : Same as top panel but for RRLs in the VVV survey.}
\label{fig:mu_vvv}
\end{center}
\end{figure}

We use $K_s$-band mean-magnitudes for T2Cs in the Galactic bulge to determine a distance to the Galactic center using the absolute calibration of PLR for T2Cs and RRLs in the MW and in the LMC. We only use short period BLH and WVR stars and apply separate calibrations based on MW and LMC T2C PLRs to the bulge data to obtain two sets of individual distances. We take an average of the two distances for each BLH+WVR type variables in $K_s$-band. Fig.~\ref{fig:mu_vvv} displays the histogram of individual distance estimates for BLH+WVR and RRLs variables. We include OGLE-IV counterparts of the RRLs from VVV survey in our analysis for a relative comparison. The RRLs sample consists of more than 20 thousand variables with high-quality light curves and their photometry will be discussed in a future publication.

\begin{figure*}
\begin{center}
\includegraphics[width=1.0\textwidth,keepaspectratio]{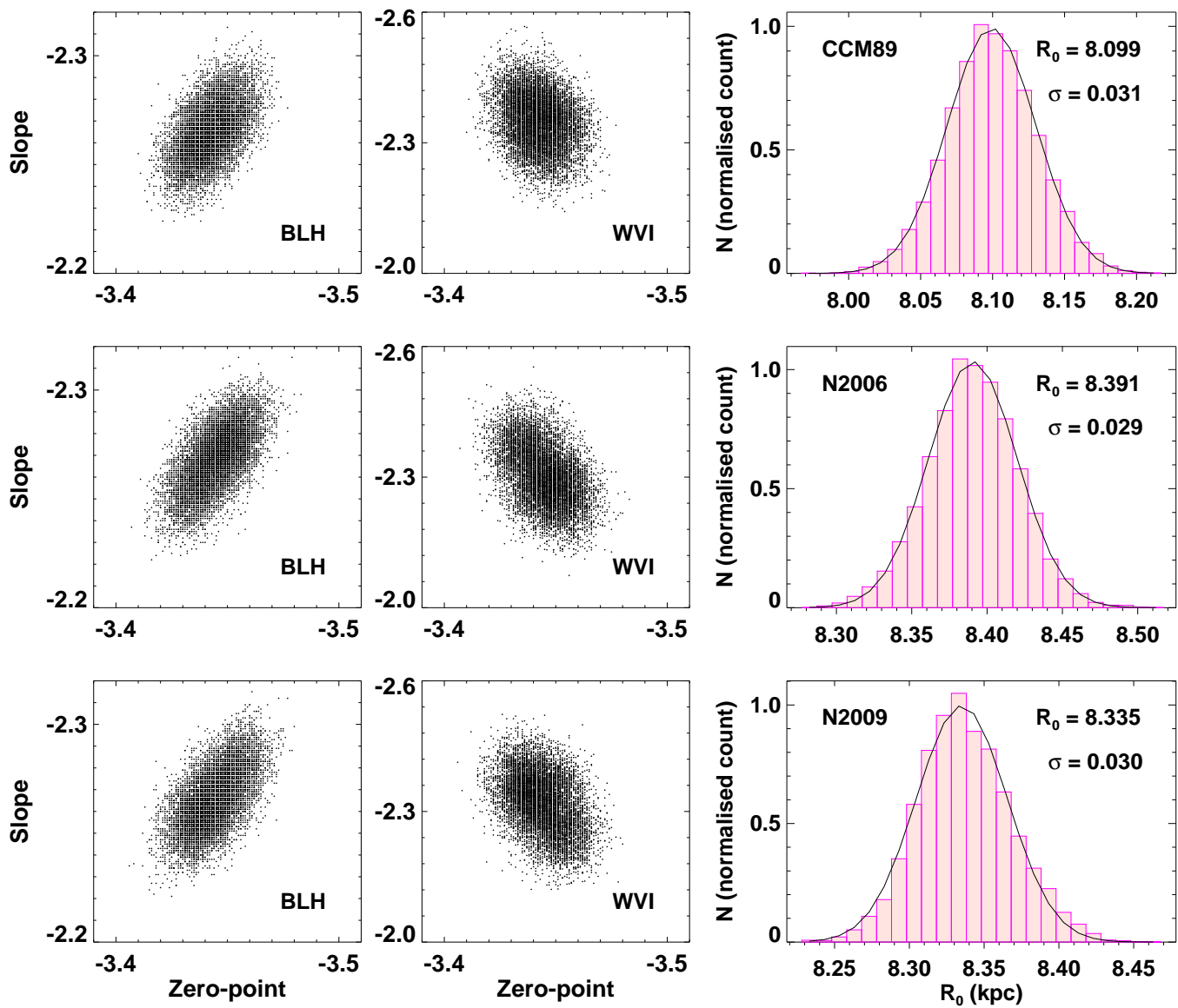}
\caption{Result of $10^4$ random realizations of parameters of the global fit to Type II (BLH+WVR) Cepheids in the globular clusters, LMC and the Galactic bulge. Left and middle panels: Slope and the zero point of global fit, Right panels: distance estimates from $10^4$ permutations. CCM89 : \citet{card89}; N2006 : \citet{nishiyama2006}; N2009 : \citet{nishiyama2009}.}
\label{fig:mu_ejk}
\end{center}
\end{figure*}

To determine the distance to the center of the population, we need to apply two geometric corrections. First, the individual distances are projected onto the Galactic plane. This is done by taking cosine of the Galactic latitude, resulting in a distance of $Rcosb$. Secondly, we need to correct the distance distribution for the ``cone-effect'' which leads to bias in distance estimates as more objects are observed at larger distances in a solid angle. This is corrected by scaling the distance distribution by $R^{-2}$. The two histogram bars in Fig.~\ref{fig:mu_vvv} represent the distances before (blue) and after (magenta) these corrections. In our sample the correction is marginal ($\sim0.5\%$) in terms of the change in the peak of the distance distribution. We note that both T2Cs and RRLs provide similar estimates for the distance to the Galactic center. The RRLs distance distribution is based on empirical calibration of T2C PLR and does not include any metallicity corrections.

In order to obtain a robust distance estimate, we use $K_s$-band mean-magnitudes for the T2Cs and apply a simultaneous fit to all BLH+WVR stars in the Galactic bulge, LMC, Galactic globular clusters and T2Cs in the solar neighbourhood. We fit a PLR in the following form -

\begin{equation}
m^0_{ij} = a_S[\log(P_{ij}) - 1] + a_L[\log(P_{ij}) - 1] + M_{K_s} + \mu_{j},
\end{equation}

\noindent where $m^0_{ij}$ is the extinction-corrected magnitude for a $i^{th}$ T2C in the $j^{th}$ system. The coefficients $a_S$ and $a_L$ represent the slopes of BLH and WVR stars and $M_{K_s}$ is the absolute magnitude in $K_s$-band for a T2C with P = 10 days. The distance modulus to the Galactic bulge is given by $\mu_{j}$. The matrix equation is solved using the chi-square minimization \citep{bhardwaj2016a}. 

\citet[][and references therein]{nishiyama2009, matsunaga2016} showed that the selection of reddening law leads to significant difference in distance estimates close to the Galactic center. Therefore, we use three different extinction-laws, $\frac{A_{K_s}}{E(J-K_s)}=0.689$ \citep{card89}, $\frac{A_{K_s}}{E(J-K_s)}=0.494$ \citep{nishiyama2006} and $\frac{A_{K_s}}{E(J-K_s)}=0.528$ \citep{nishiyama2009}, in our analysis.  

\begin{table}
\begin{minipage}{1.0\hsize}
\begin{center}
\caption{Parameters of the global fit.}
\label{table:global}
\begin{tabular}{cccc}
\hline
\hline
 & CCM89 & N2006 & N2009 \\
\hline
$a_S$&     -2.261$\pm$0.013     &    -2.266$\pm$0.014     &    -2.261$\pm$0.014     \\
$a_L$&     -2.354$\pm$0.062     &    -2.299$\pm$0.068     &    -2.328$\pm$0.066     \\
$M_{K_s}$&     -3.438$\pm$0.010     &    -3.441$\pm$0.011     &    -3.439$\pm$0.010     \\
$\mu$&     14.552$\pm$0.008     &    14.630$\pm$0.008     &    14.616$\pm$0.008     \\
$R_0$&	    8.099$\pm$0.031     &     8.391$\pm$0.029     &    8.335$\pm$0.030 \\
\hline
\end{tabular}
\end{center}
{\footnotesize {\bf Notes:} CCM89 : \citet{card89}; N2006 : \citet{nishiyama2006}; N2009 : \citet{nishiyama2009}.}
\end{minipage}
\end{table}

\begin{figure*}
\begin{center}
\includegraphics[width=1.0\textwidth,keepaspectratio]{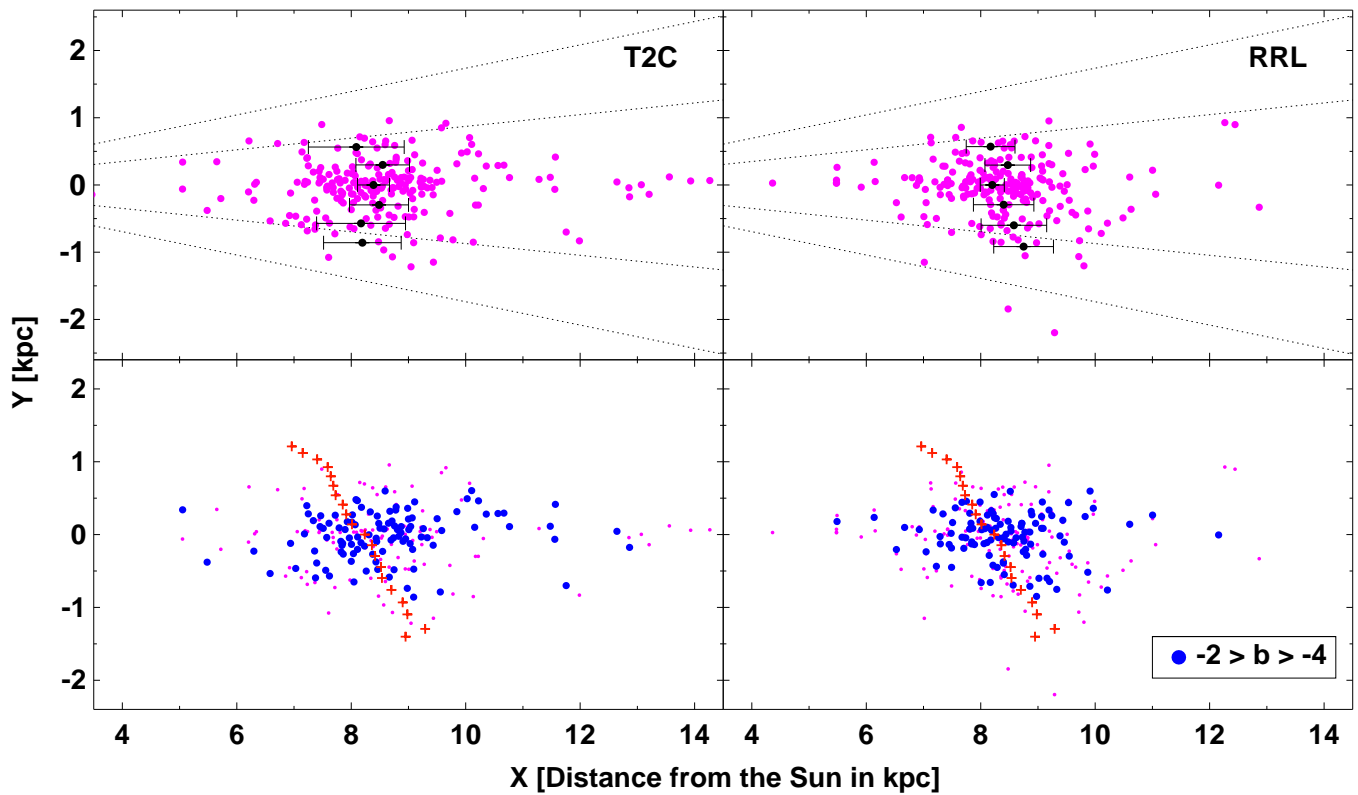}
\caption{Top panels : The Spatial distribution of T2Cs (left column) and RRLs (right column) projected on to the Galactic plane. RRLs are drawn for each representative T2C from the parent sample of RRLs in the VVV survey. The black circles represent projected mean line-of-sight distances in different longitude bins and error bars display $3\sigma$ standard deviation of the mean. The dotted lines represent the line-of-sight corresponding to $l = \pm5$ and $l = \pm10$. Bottom panels: The spatial distribution of T2Cs and RRLs in the latitude stripe $-2 > b > -4$ (large blue dots) and outside these latitudes (small magenta dots) are compared with the mean red clump giants' distances (red plus symbols). The mean distances to red clump giants are derived from peaks of Gaussian distributions in each 1 sq. deg. field for latitudes in the range $-2 > b > -4$ \citep{gonzalez2012,valenti2016}, and assuming a mean absolute magnitude $M_{K} = -1.61\pm0.015$ \citep{laney2012}.}
\label{fig:spatial}
\end{center}
\end{figure*}

We carry out Monte-Carlo simulations to create $10^4$ random-realizations of the global fit. We estimate the coefficients of the global fit for each permutation and fit a Gaussian function to the histograms to determine a mean value with their statistical uncertainties. Fig.~\ref{fig:mu_ejk} displays the variation of $10^4$ random realizations of the slopes and the zero point (left and middle panel) and the histogram of distance estimates (right panel) for three different extinction laws. The coefficients of the global fit are given in Table~\ref{table:global}. The slope of WVR type variables shows a greater variation as compared to BLH type variables. As expected, due to similar extinction coefficients, the difference between distance estimates from the two Nishiyama's extinction laws is not significant, but the distance to Galactic center differs significantly when compared to the distance obtained using \citet{card89}. We adopt the distance to the Galactic center, $R_0 = 8.34\pm0.03$~kpc, obtained using the reddening law of \citet{nishiyama2009}. This value is close to the value obtained using \citet{nishiyama2006} reddening law, but more consistent with recent distance estimates from other distance indicators, such as, RRLs from VVV survey, $R_0 = 8.33\pm0.05(\textrm{stat.})\pm0.14(\textrm{syst.})$~kpc \citep{dekeny2013} and from OGLE-IV, $R_0 = 8.27\pm0.01(\textrm{stat.})\pm0.40(\textrm{syst.})$~kpc \citep{piet2015}. Our distance to the Galactic center is in excellent agreement with recommended distance, $R_0 = 8.3\pm0.2(\textrm{stat.})\pm0.4(\textrm{syst.})$~kpc, by \citet{degrijs2016}. We note that these distance estimates are also consistent with results based on studies of stellar orbits in the Galactic center and distance to central black hole \citep[$R_0=8.32\pm0.07(\textrm{stat.})\pm0.14(\textrm{syst.})$~kpc, ][and references therein]{genzel2010, gillessen2017}. We adopt a conservative approach for the systematic uncertainty on the distance to the Galactic center and include the error on the photometry (median $\sim0.08$~mag), the error on the $A_{K_s}$ extinction values (median $\sim0.07$~mag), uncertainty in the zero-point of the calibrated PLRs (0.02~mag, inverse weighted variance resulting from two independent calibrations), which amounts to $\sim0.11$~mag ($0.41$~kpc). 

\section{The Spatial distribution of Population II Cepheids} 
\label{sec:spatial}

The distribution of old metal-poor stellar populations unveil an axisymmetric component of the bulge while metal-rich red-clump giants show an elongated distribution that trace the bar \citep{gonzalez2011, zoccali2017}. Metal-poor RRLs, Miras and red-clump giants display an axisymmetric and spheroidal structure of the bulge \citep{dekeny2013, catchpole2016, zoccali2017}. RRLs also show a triaxial ellipsoid shape with OGLE-IV data \citep{piet2015}. We look at the spatial distribution of T2Cs projected on to the Galactic plane. We also select a representative RRL within $1'$ radius of each T2C position from the distribution of RRLs shown in Fig.~\ref{fig:mu_vvv} and adopt a median distance. 

The top panels of Fig.~\ref{fig:spatial} displays the spatial distribution of T2Cs, projected onto the Galactic plane. The T2Cs sample shows a homogeneous distribution with majority of distance estimates falling within 6-10 $kpc$. The corresponding RRL distribution is more spherically symmetric. The T2Cs distance distribution shows more stars at longer distances as compared to RRLs subsample which is slightly elongated towards us. We also overplot mean line-of-sight distances in different longitude bins for T2Cs and RRLs. Neither population provides evidence of an inclined bar or X-shaped structure as traced by the metal-rich red-clump giants. We note that majority of T2Cs in the current sample are located along $b=-2$ to $b=-4$ and their spatial distribution is shown in bottom panel. The distribution along these latitudes confirms no-barred distribution with current T2C and RRL samples in the bulge. 

We also performed two-sided KS-test to compare the distance distributions of T2Cs and RRLs from VVV survey. We find that both these populations have similar radial distributions in most spatial bins. However, the number density of T2Cs is very limited at present and the results of test-statistics are heavily influenced by small number statistics and the choice of bin-size. In near-future, with more data from OGLE-IV and VVV, T2Cs can become additional important tracers of the bulge metal-poor old populations, probing its structure and formation.

\begin{figure}
\begin{center}
\includegraphics[width=0.5\textwidth,keepaspectratio]{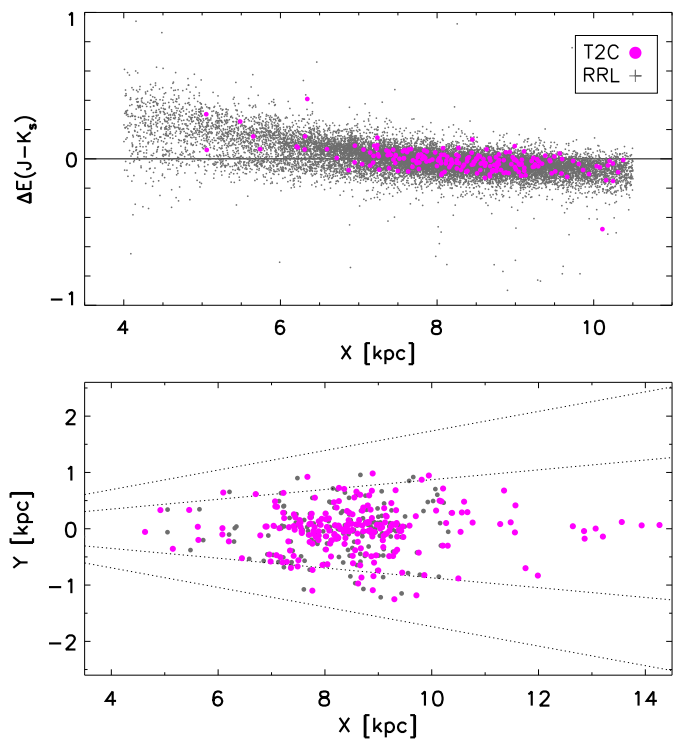}
\caption{Top panel: The difference in $E(J-K_s)$ values for T2Cs and RRLs obtained from extinction maps based on red clump giants \citep{gonzalez2012} and 3-D extinction map of \citet{schul2014} as a function of distance. Bottom panel : The spatial distribution of T2Cs before (grey) and after (magenta) the extinction correction using 3-D map.}
\label{fig:del_ejk}
\end{center}
\end{figure}

We recall that our T2C and RRL distance estimates are based on $K_s$-band mean-magnitudes from VVV survey.
We do not use Wesenheit relations to get the distance estimates as the time-series observations are available 
only in $K_s$-band. Furthermore, given that $K_s$-band PLR has least dispersion (see Fig.~\ref{fig:pl_cal} and 
Table~\ref{table:gb_pl} \& \ref{table:comp_pl}), we deem it is better to apply extinction correction to $K_s$-band 
PLR adopting external reddening values than constructing a Wesenheit with random-phase corrected $JH$ magnitudes.
The extinction correction is applied using \citet{nishiyama2009} reddening law and $E(J-K_s)$ color excess values from the extinction maps of \citet{gonzalez2011, gonzalez2012}, which are derived by comparing the mean $E(J-K_s)$ color of red clump giants in small subfields of $2'\times2'$ to $6'\times6'$ with the color of red clumps in Baade’s window \citep[see,][for details]{gonzalez2012}. The color excess for each T2C and RRL is obtained by inserting their longitude/latitude to Bulge Extinction And Metallicity Calculator (BEAM){\footnote{\url{http://mill.astro.puc.cl/BEAM/calculator.php}}} and adopting a resolution of $2'$.

Given that the 2-D BEAM extinction map assumes all extinction is at the location of the tracer (red clumps, which are mostly located in the bulge/bar), while our targets are located along the line-of-sight at different distances with respect to the bar, we compare the BEAM $E(J-K_s)$ values with $E(J-K_s)$ from 3-D extinction map of \citet{schul2014}. The 3-D extinction maps are based on the VVV data and temperature-color relation for M giants and the distance-colour relations. The stellar population synthesis models were used to offset observed and intrinsic colors and obtain extinction maps with two spatial and one distance dimension.

We compare the distance estimates for each T2C and RRL at a given longitude and latitude and obtain $E(J-K_s)$ color excess from 3-D maps. The difference in $E(J-K_s)$ color excess values as a function of distance is shown in the top panel of Fig.~\ref{fig:del_ejk}. The difference increases for the line-of-sight distances towards us and the offset is $\sim-0.05$~mag for T2Cs and $\sim-0.03$~mag for RRLs close to the Galactic center. We also correct the extinction values from \citet{gonzalez2012} and corresponding distance estimates by iteratively computing the difference in extinction with respect to 3-D maps. The initial $E(J-K_s)$ values are corrected for offset and the distances are redetermined in each iteration. The procedure is repeated until extinction or distance estimates converge. 

Bottom panel of Fig.~\ref{fig:del_ejk} shows the spatial distribution of T2Cs before (grey) and after (magenta) the extinction correction from 3-D maps. At present, the accuracy of the distance dimension of 3-D maps is limited to the bin size of 0.5~kpc (upto 10.5~kpc) and therefore, the correction in extinction converges typically in second or third iteration. Thus, the difference in resulting distance distribution is very small and not statistically significant. Median difference in extinction is $\sim0.01$ and the distance to the Galactic center changes only marginally ($\sim10^{-3}$) if we use corrected extinctions from 3D maps.

\section{Conclusions}
\label{sec:discuss}

We summarize our results as follows :

\begin{itemize}

\item{We present a catalogue with mean $VIJHK$ magnitudes, periods, reddening and subclass classifications for 264 Type II Cepheids in the Galactic bulge by matching the VVV near-infrared observations with optical data from OGLE-III. The sample consists of various subtypes (BL Herculis, W Virginis and RV Tauris) with on an average of 50 $K_s$ measurements per light-curve.}\\

\item{We use random-phase corrected $JH$ magnitudes and $K_s$ mean-magnitudes to derive period-luminosity relations for Type II Cepheids. The long-period RV Tauris stars are affected by saturation and therefore a sample of BL Herculis and W Virginis stars is used in our distance analysis. The period-luminosity relation in $K_s$-band, ${K_s} = -2.189(0.056)~[\log(P) - 1] + 11.187(0.032)$, is found to be consistent with published work for the LMC variables.}\\

\item{We apply a global fit to the Galactic bulge, LMC and Galactic globular cluster Type II Cepheid data in $K_s$-band, together with calibrated absolute magnitudes for Type II Cepheids and RR Lyrae with {\it Gaia} and {\it Hubble Space Telescope} parallaxes, to determine a distance to the Galactic center, $R_0 = 8.34\pm0.03(\textrm{stat.})\pm0.41(\textrm{syst.})$~kpc. Our results are in a very good agreement with published distance measurements based on Type II Cepheids, RRLs \citep{gmat2008, dekeny2013, piet2015} and the recommended distance by \citet{degrijs2016}. Adopting a different extinction law amounts to a difference of $^{+0.05}_{-0.25}$ in the final distance to Galactic center.} \\

\item{We also investigated the spatial distribution of Type II Cepheids in the Galactic bulge. We compared their distribution with well-studied most abundant tracers in the bulge, such as, RR Lyrae and red-clump giants. We find that Type II Cepheids display a non-barred distribution, similar to other metal-poor bulge tracers, RR Lyrae \citep{dekeny2013} and red-clump giants \citep{zoccali2017}.
This result requires further validation with a larger sample of T2Cs that are uniformly distributed, in particular at low latitudes.} \\

\item We test the individual distance estimates and extinction values for Type II Cepheids in the Galactic bulge by taking into account the 3-D distribution of dust in the Milky Way. At present, considering the discretization of 3-D maps along distance dimension in steps of 0.5~kpc, the correction of individual distances is not significant.\\

\end{itemize}

\section*{Acknowledgments}
\label{sec:ackno}
AB is thankful to the Council of Scientific and Industrial Research, New Delhi, for a Senior Research Fellowship 
(SRF, 09/045(1296)/2013-EMR-I). We acknowledge the use of data from the ESO Public Survey program ID 179.B-2002 taken with the VISTA telescope. DM, MR, EV, MZ and OG acknowledge hospitality of the Aspen Center for Physics, where this work was initiated. The Aspen Center for Physics is supported by National Science Foundation grant PHY-1066293. DM and MZ were partially supported by a grant from the simons Foundation, during their stay in Aspen and gratefully acknowledge support by FONDECYT Regular grant No. 1130196 and 1150345, by the Ministry of Economy, Development, and Tourism's Millennium Science Initiative through grant IC120009, awarded to The Millennium Institute of Astrophysics (MAS) and by the BASAL-CATA Center for Astrophysics and Associated
Technologies PFB-06.


\bibliographystyle{aa}
\bibliography{popII_cep}

\end{document}